# Critical Considerations for Developing MIS for NGOs

By

**Kailash Chandra Dash**

Bhubaneswar, India kailash@rcdcindia.org

**Umakant Mishra**

Bangalore, India umakant@trizsite.tk

## Contents

## Abstract

Although Information Systems and Information Technology (IS & IT) has become a major driving force for many of the current day organizations, the NGOs have not been able to utilize the benefits up to a satisfactory level. Most organizations use standard office tools to manage huge amount for field data and never feel the need for a central repository of data. While many people argue that an NGO should not spend too much money on information management, it is a fact that organizing the information requires more of a mindset and an organized behavior than a huge financial investment.

## 1. Introduction

An NGO[1] (Non Governmental Organization) generally works for social causes like the betterment of society, empowerment of the weaker sections, conservation of environment and natural resources etc. Often the activity of an NGO looks simple and low profile because there is no great reward for a true achievement and no punishment for a failure either. But achieving the goals of an NGO is no easier than fulfilling the toughest goals of any commercial organization. Often the NGOs work on issues on which commercial organizations never work and the government fails to proceed.

A Management Information System (MIS) is a systematic organization and presentation of information that is generally required by the management of an organization for taking efficient and effective decisions for the organization[2]. As the goals and structure of every organization is different their MIS needs are also different. The MIS required for an NGO is different from other organizations as there is no profit intention or expectation from the NGO activities. The core data of NGO MIS must concentrate on its core activities and objectives (such as, social or developmental) for which the NGO is working and existing.

## 2. Need for MIS in NGOs

The days are gone when few individuals were working for the poor and needy out of their benevolence. Now the developmental activities are being carried out by larger groups of professionals with clear objectives and strategies. Although there are plenty of funds available for developmental activities from governments, corporate houses, individuals and societies, the competition between the NGOs has grown significantly. The NGOs are being judged by their capability of implementing projects. The funding agencies evaluate the strength of an NGO through its past activities and success stories. The funding agencies extend their financial support only after satisfactory evaluation of NGO performances and cost of their services.

Under such circumstances an NGO has no option but to improve its efficiency through various means. It has to prove that it not only has the expertise on specific developmental issues but also has the capability of managing larger projects in the right way to achieve the desired goals. This is where a Management Information System (MIS) plays a great role. Without a proper information system the data will be scattered in bits and pieces of papers and computers with different people. There will be no way to measure and compare different field activities carried out in different remote locations. Every time there is a need to present its activities the NGO has to collect information from its field offices, project managers, and lost/forgotten sources. Often employees leave taking the data/information with them. The MIS solves this problem by keeping all the important information in one central repository and making the information available to the management.

---

[1] An NGO is also known as a Voluntary Organization or Non-Profit Organization
[2] Umakant Mishra, Introduction to Management Information System, Cornell University Library, http://arxiv.org/abs/1308.1797

Today's NGOs are often spread over a vast geographical area handling a large number of diverse projects. Data needs may be vastly different for each project. For example data requirement for a governance project will vary drastically from an agriculture based project. There will be both macro and micro requirements depending on the nature of the program. There are also cross cutting issues that require consolidation of data from various projects and activities. Moreover NGOs do not work with the same donors for a long period of time. Thus with change of donors there will be changes in the requirements of data and reporting even if the same program is continued but with a different donor. Data requirements may be both internal and external. While internal data is generated from implementation activities, external data from Government and other agencies is required for comparisons, advocacy purposes, planning and policy making.

Data generated will be used not only by the program team but also by other staff who need particular data from other programs. Those involved in preparation of proposals and annual reports will need access to the entire database. A high degree of flexibility is therefore required in any MIS that caters to the needs of NGOs.

This flexibility should be available both in capturing data as well as in the formats used to store the data. A dynamic system is required which cannot be satisfied using Word or Excel formats. The MIS system must be user friendly to the extent that both basic and query formats can be changed without disturbing the database in anyway. The user also must be able to create new formats and make changes to existing formats either by making changes to the original structure or by importing existing data into the new formats.

The data requirements are not limited to program requirements. The finance and administration teams are in acute need for databases for finance, administration and for meeting legal/tax requirements. As the finance teams usually have separate accounting software, there is need to examine if an MIS system can be developed which can incorporate the final reports of the accounting software. The administration team needs to keep track of files, assets, leave status, and other basic details of each staff. The program teams also need financial and administrative data to know about the budgetary aspects, HR and assets needs of the program. As there is high turnover of employees there is the need to capture assets assigned to staff and their leave and other status such that recoveries can be made promptly and salaries paid.

The MIS system not only needs to capture data but also documents, both completed and as work in progress. Old important documents available only in print must also be scanned and uploaded. Similarly the progress of the programs can also be ascertained by using the MIS. For this separate work spaces for each staff can be created where they can work as well as store their daily diaries, plans and activity reports as well as unfinished reports and documents. Once finished, the completed reports can then travel to the work space of senior staff that can verify the reports and then add them to the central repository for onward dispatch to intended recipients as well as for storing. The work spaces can be clubbed team wise as modules but yet be accessible to members of other teams who have been authorized by the administrator.

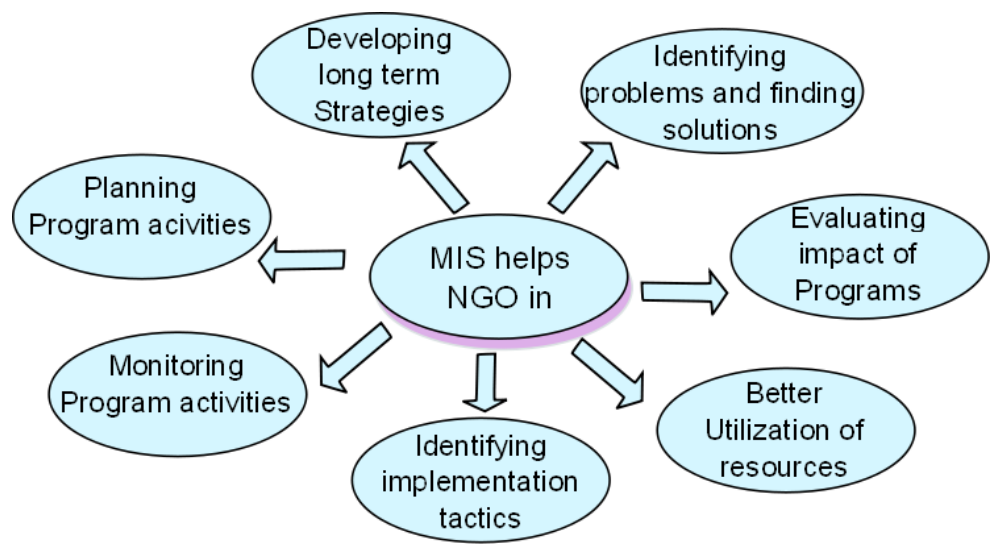

Many people feel that if the people of the organization are good then the organization will do well. This is true, but is not the complete picture as a good manager can become inefficient without the required information. The program director must have immediate access to the latest data of each activity implemented in every project. The availability of right data at the right time increases the efficiency of the program directors to take quick and effective decisions. Some of the benefits of having an MIS are:

**Benefits of having an MIS**

- ⇨ An MIS facilitates easy access to all the important data of the organization to the people in the management.
- ⇨ The MIS often keeps the data organized in a central location and makes the information available for quick analysis and presentation.
- ⇨ Quick availability of up-to-date information facilitates monitoring and self-correction and improves controllability and operational efficiency of the organization.
- ⇨ An efficient MIS helps strategic and long-term decision-making and improves the overall productivity of the organization.
- ⇨ An MIS can eliminate the incompatibilities and inconsistencies found in different formats/procedures used in different projects and/or at different times and brings compatibility and standardization of formats/procedures
- ⇨ The accountability of the NGO is addressed by publishing its activities and achievements to the donors, funding agencies, government and to the public.
- ⇨ Higher efficiency and better presentability increases the reputation and credibility of the NGO.
- ⇨ A good MIS will obviously improve operational efficiency, employee productivity, project efficiency and funding agency/donors' satisfaction.
- ⇨ The NGO gains competitive advantage to attract more program funds, experienced staff and appropriate projects.

## 3. Considerations for building an effective MIS

Most of the NGO information are generally presented in simple lists and tables. Hence most NGOs manage their data by using word processors and spreadsheets. However we also explained the drawbacks of managing data in spreadsheets. For example, there might be multiple copies of the same reports creating confusion about which is final and which is not. Besides some reports might get lost and regenerated by the program officers later when required. Data in spreadsheets remain with specific individuals and may vanish when people make operational mistakes or when people leave the organization. It is necessary to move from the concept of personal data to an organizational database, from a spreadsheet based data management to a centralized database. Let us examine the key considerations of building an effective MIS.

### Understanding the organizational need

Unlike commercial organizations, the NGOs don't go after profit. The NGOs exist for a purpose and a goal. The MIS should address to its core activities and objectives. The implementation may be different in different NGOs depending on the structure of the organization, reporting procedure, core activities, nature of projects, hierarchy of data urgency and importance etc. It is also important to consider the negative sides, e.g., difficulties of MIS development, data collection, data entry etc.

### Organizational commitment

The most important thing required to build an MIS is an organizational commitment. Like you have a budget for trainings, library and procurements etc. so you need to have a budget for MIS too (whether small or big). Often the executive director has to spend enough time in understanding the MIS structure, requirements, and capabilities etc. to utilize the MIS in the best possible way to get the strategic advantage.

### Balancing the extremes

Every design of MIS has its scope and limitations. While adding too much operational data can make the MIS bulky and clumsy, non-inclusion of crucial data makes the MIS weak and ineffective. Hence there should be a perfect balance between the volume and importance of data. There should be a harmony between the executive director and the MIS expert to match the requirements with the boundaries.

### Integrating NGO activities with Technology

While there are a large number of ready-made software available out of the box for many regular activities there is no such ready made software for MIS. Although there is software available for analyzing financial data or managing customer relationship etc. there is no such ready made tool to manage the NGO data or data on developmental projects. It is important to consider the following issues while investing on custom built software.

- ⇨ Developing custom built software often becomes a long process and expensive.
- ⇨ It is important to select a consultant/ firm who have proven experience on developing similar MIS tools.
- ⇨ The organization should provide sample data on whatever project required. Feeding real life data can assess the functionality of the system.
- ⇨ It is possible to develop different modules of MIS in phases. One may decide to develop and implement modules phase by phase. But it is not possible to develop different modules by different developers as they will fail to collaborate.

- Training on MIS may be important before implementing a new MIS project. The managers/officers responsible for entering data should be explained about how to enter the data.

## Problem of Harmonizing NGO terminologies with IT

Generally NGO managers and software developers talk in different languages. While NGO managers have a clear concept of their operations they many fail to explain those properly to the MIS developer to build the right database and interface. The MIS developer tries to analyze the NGO requirements and tries to build a system that is best understood through the analysis. Any mistake in the requirement analysis may lead to:

- Inclusion of unimportant data items thereby making the MIS clumsy and increasing the burden of data entry.
- Non-inclusion of important data items thereby making the MIS weak and useless.

Hence, the MIS needs should be clearly explained to the consultant and the consultant should produce system analysis and design documents to ensure that the needs are properly explained and understood.

## Customization and Flexibility

Often an NGO works with multiple projects having some similarities and some differences between them. If we emphasize on the differences then there will be too much work for defining each project activity uniquely. This will lead to incompatibility with future projects and will require continuous modification and dependency on the MIS developer. However, on the other hand developing a generic product for accepting any kind of project parameters can require very complex design and programming logic.

## Possible Problems with MIS developer

When the MIS application is to be developed by an outsider there are many possible problems. In most cases the application developer will not agree to give you the source code. This is a complicated situation.

- In the first case there will be a dependency on the application developer for all future changes.
- Secondly if the application developer puts a key to be renewed in the web application then there will be compulsory dependency on the developer.
- Thirdly, if the application developer goes out of market then there is no way to modify the program.

One solution to the above problem is to buy the copyrights on the source code from the developer. This will help you to modify as and when required and even to sell the program to other organizations.

Another solution is to get the source code and give a written commitment to the developer stating to maintain due confidentiality and modify only when absolutely necessary.

## The difficulties of data entry

MIS software is lifeless without its data. The success of an MIS project depends upon the regular feeding of its data. But it is possible that some officers may not like to enter data into the MIS because of various reasons.

- ⇨ The data entry forms may be deterrent because of their complicacy and lack of flexibility. For example a "tour entry form" may not save without entering "date of visit" and the person entering data may not have the "date of visit" available at the time of entry.

- ⇨ Often people dislike data entry forms as they require the data to be in specific format. For example, a date field (say, date of commencement) will not accept "Jun 13" as it is not a valid date format. People prefer word processors and spreadsheets as they can enter anything in any format. **Solution** - It is the challenge of the MIS developer to ensure that the MIS entry forms should be as easy and flexible as entering data in a spreadsheet.

- ⇨ Sometimes the people engaged in fieldwork are more comfortable in verbal expression than interacting with computer. **Possible solution**- the MIS should have free style forms to enter whatever the person wants to enter.

- ⇨ Even if the entry screen is easy and user friendly some program officers still may not like it. One of the hidden reasons is that they cannot change the previous months data so easily/ efficiently as they can do with it in spreadsheets. **Solution**- this difficulty can be overcome by developing a mindset of sincerity in terms of giving monthly figures.

## Web based or LAN based technology

There is no single way of developing an MIS. An MIS can be LAN based and can be web based too. **LAN based**- if you have most of the management staff located in the head office, then it will be good to implement a LAN based MIS. LAN based technologies can be fast to develop by using Rapid Application Development tools. **Advantages**- easy to manage, faster access, no dependency on external server or internet connection. **Disadvantages**- Cannot be accessed from outside. **Internet based**- **If** you have offices at different locations or people in management are mostly traveling then it is advisable. **Advantages**- You can access / update data from different locations/ anywhere in world. **Disadvantages**- difficult to program and maintain, cannot be accessed without Internet. **Hybrid**- Maintain a LAN based MIS and update monthly reports on the Internet.

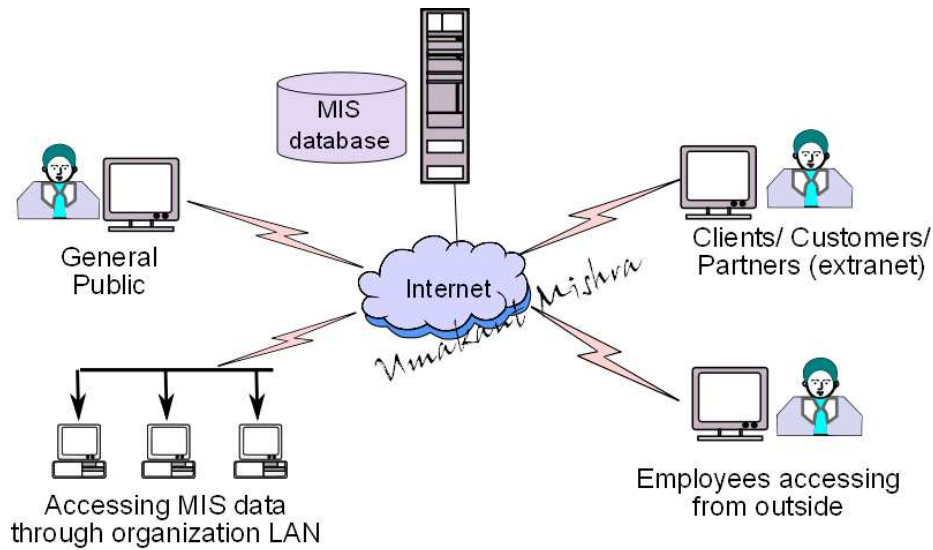

Implementing MIS through Internet

When you choose an Internet based MIS the data is supposed to be placed on a server which is not within the client's office. If the data is very confidential then the organization has to take adequate security measures to maintain the confidentiality of the data.

**Data security**

The MIS data of any organization is not only critical but also (at least part of it is) confidential. No organization would like to see its private data being viewed by unauthorized people or confidential data being exposed to the public. Hence it is important to implement proper security measures to ensure that the valuable organizational data is protected from being misused by wrong people. The MIS developer may implement proper access rights[3] to ensure that right portion of the MIS data is available only to the right people.

**Role of MIS Manager**

Although the MIS database and software can be build by outside agencies somebody from the organization, whether MIS manager or administrator or director, should keep close eyes on the project. He has to play a vital role in interfacing various groups, training the new users, feeding the master data, assigning access rights, keeping data backup, ensuring data security and confidentiality etc. The person should be highly responsible, should have good knowledge on IT and good understanding of NGO program implementation.

---

[3] Umakant Mishra, How to Implement Access Rights in an MIS Project, http://ssrn.com/abstract=2315570

## 4. Characteristics of a good MIS

An ideal MIS system should have all advantages and no drawbacks[4]. It should be easy to develop, easy to maintain and easy to use. It should be cheap or free of cost. It should provide all the data required by the management.

- ⇨ The first and foremost characteristic of a good MIS is that it should be able to provide all the important data to the management, nothing more and nothing less. For example, it may only provide monthly reports or summary reports which are useful for management decisions.

- ⇨ A good MIS should work well with the existing infrastructure. It should not require additional hardware, operating system or other software to be purchased. Even if some software is required it should be available free[5] (or at nominal cost). Besides it should be easy to work with and should not require any special skills or special trainings to work with.

- ⇨ It may be a good idea to integrate MIS and intranet. An intranet may consist of some document sharing and collaboration tools. The MIS also does a similar job although with different data and for a different purpose. Harnessing one on top of the other may yield better results.

- ⇨ An MIS is generally designed for a long term say more than 5 years. That means a well designed MIS should require no modifications on a year to year basis.

- ⇨ Apart for reports required by the management, the MIS may generate some reports which are required by the donors, government or the public.

- ⇨ The MIS should be flexible to include more projects and more data types as and when required in future without needing to change its database structure and program codes.

- ⇨ Ideally the same MIS engine should work for all types of projects, whether long or short, whether on income generation or on water-sanitation or on animal rights. The same engine should work for all the projects of past, present and future. There should be no need to call the MIS developer to accommodate a new project which is different from the previous.

---

[4] Umakant Mishra, Introduction to the Concept of Ideality in TRIZ, Oct 2007
http://papers.ssrn.com/abstract=2273178

[5] Umakant Mishra, The Concept of Resources in TRIZ, TRIZsite Journal, Aug 2007
http://papers.ssrn.com/abstract=2212093

To explain the above issue, the MIS should accept user defined fields. As every project has different activities and different parameters to be monitored the MIS should allow feeding new monitoring parameters (fields) and allow feeding data against those monitoring parameters.

⇨ The MIS should maintain adequate security. The MIS data is expected to be available only to the authorized users. The MIS may implement different levels of access to allow or disallow different people for different types of data. The data entry forms should be accessible only to specific authorized people.

Different levels of management may be given different levels of access to different data in MIS. While more people will have only read access only a few will have write (entry) access.

⇨ There should be no dependency on the MIS developer. One option is to get the source code from the MIS developer. However, this is not always an easy case. The NGO must deal on high moral grounds and give written commitment on keeping the source code confidential only to be used in emergency.

## 5. Summary and Conclusion

Although the Information systems have become a major driving force for many of the current day organizations, the NGOs have not been able to utilize the benefits of these systems at the desired level. Most NGOs use standard office tools to manage huge amount of field data collected month by month for every project. Although many people argue that an NGO should not spend too much money on information management, it is a fact that organizing the information requires more of a mindset and an organized behavior than a financial investment.

While an MIS should provide all useful information that is required by the management the management should also have a commitment for MIS. The MIS manager should ensure that the rights are assigned to the right people to maintain data security. The MIS should be user friendly and data entry should be easy and interesting so that people would love to enter their data into the MIS. The design of the MIS should be flexible to accept any type of project undertaken by the NGO and should work flawlessly for long term without needing to be modified on getting new types of projects.

## About the Authors:

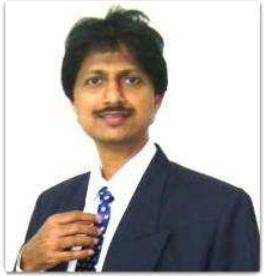

Umakant last worked as Chief Technology Officer in CREAX Information Technologies (Bangalore) and before that IS/IT manager for ActionAid India (Bangalore). He has authored a few books "TRIZ Principles for Information Technology", "Improving Graphical User Interface using TRIZ", "Using TRIZ for Anti-Virus Development" etc. and authored more than 100 articles in TRIZsite Journal, SSRN eLibrary http://ssrn.com/author=646786 etc. More about Umakant is at http://umakant.trizsite.tk.

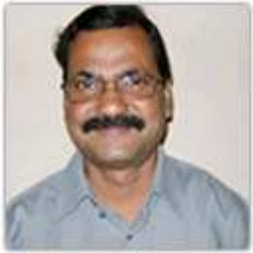

Kailash Chandra Dash is currently working as Executive Director at Regional Center for Development Cooperation (RCDC) at Bhubaneswar (http://rcdcindia.org). He is coordinating about 10 developmental projects in various regions of Odisha. Before that he has worked in senior positions in organizations like Danish International Development Assistance (DANIDA), Indo Global Social Service Society (IGSSS), Indian Red Cross Society etc.